\documentclass[useAMS,usenatbib]{mnras}

\usepackage[fleqn]{amsmath}
\usepackage{amsfonts,graphicx,txfonts,fleqn,color,tabularx,verbatim,gensymb}
\usepackage{makecell}
\renewcommand*{\vec}[1] {\boldsymbol{#1}}
\newcommand*  {\kms}   {\mbox{km\,s$^{-1}$}}
\newcommand* {\auq}{\mathrm{au}^{-3}}

\title[Capturing interstellar comets]{Capture of interstellar objects: a source of long-period comets}
\author[Hands, Dehnen]{T.~O.~Hands$^1$\thanks{email:
tomhands@physik.uzh.ch}, W.~Dehnen$^{2,3}$
\\$^1$Institut f\"ur Computergest\"utzte Wissenschaften, Universit\"at Z\"urich, Winterthurerstrasse 190, 8057 Z\"urich, Switzerland
\\$^2$Department of Physics \& Astronomy, University of Leicester, University Road, Leicester, LE1 7RH, UK
\\$^3$Universit\"ats-Sternwarte der Ludwig-Maximilians-Universit\"at, Scheinerstrasse 1, M\"unchen D-81679, Germany
}
\begin{document}

\pagerange{\pageref{firstpage}--\pageref{lastpage}} \pubyear{2019}

\maketitle

\label{firstpage}

\begin{abstract}
We simulate the passage through the Sun-Jupiter system of interstellar objects (ISOs) similar to 1I/`Oumuamua or 2I/Borisov. Capture of such objects is rare and overwhelmingly from low incoming speeds onto orbits akin to those of known long-period comets. This suggests that some of these comets could be of extra-solar origin, in particular inactive ones. Assuming ISOs follow the local stellar velocity distribution, we infer a volume capture rate of $0.051\,\mathrm{au}^3 \mathrm{yr}^{-1}$. Current estimates for orbital lifetimes and space densities then imply steady-state captured populations of $\sim10^2$ comets and $\sim10^5$ `Oumuamua-like rocks, of which 0.033\% are within 6\,au at any time.
\end{abstract}


\section{Introduction}
Comets have fascinated humanity for centuries. These exotic objects present what many believe to be an immaculate sample of the early Solar system. Classically, comets have been considered objects that were scattered by the giant planets from a trans-Neptunian disc into the outer reaches of the Solar system, forming the \cite{Oort1950} cloud, a reservoir for comets. Passing stars may then scatter these comets back into the inner Solar system where we observe them as long-period ($P > 200$yr) comets \citep[see][for a review]{Dones2004}. This remains the most popular theory for the origin of long-period comets (LPCs), with increasingly detailed modelling of the formation and subsequent perturbation of the Oort cloud matching the observed population of such comets closely \citep[see e.g.,][]{Vokrouhlicky2019}. 

The discovery of our first known interstellar visitor 1I/`Oumuamua \citep{Meech2017,ISSI2019} raised the intriguing possibility that we might one day be able to understand the planet formation environment around alien stars by studying asteroids or comets ejected from them. The nature of `Oumuamua was immediately questioned with some studies suggesting that the object displayed minor levels of cometary activity \citep{Micheli2018}, some suggesting it did not \citep{Rafikov2018} and some setting relatively low upper limits on any potential activity \citep{Trilling2018}. Unfortunately, the transient nature of `Oumuamua means that it is impossible to gather further data, and curious parties were forced to await the chance detection of another Interstellar Object (ISO). Fortunately this opportunity presented itself again with the recent discovery of 2I/Borisov \citep{Tomatic2019}. This object has a larger velocity than `Oumuamua and much more obvious cometary activity \citep[see e.g.,][]{Guzik2019,Fitzsimmons2019}. It was also discovered relatively early on its approach to the Solar system, meaning we can expect further observations in the coming months. There are of course many more known objects with hyperbolic orbits that have passed the Sun, and although most of them are thought to be ejected native comets, \cite{deLaFuente2018} highlighted 8 such objects that are potentially of interstellar origin. It is of course also possible that an interloper interacts with a  planet and becomes bound to the Solar system, in which case it might be possible to study rocky material from another stellar system in finer detail.

The study of the capture of ISOs and their incorporation into the Solar system dates back at least as far as \cite{Valtonen1982}, who performed simulations of ISO capture by the Sun-Jupiter system and concluded that this process is too unlikely to significantly contribute to the Oort cloud. Prior to this, several authors \citep{Havnes1969, Everhart1972, Everhart1973} had suggested that Jupiter could alter the orbits of near-parabolic ($e \simeq 1$) long-period comets, bringing them into the inner Solar system as short-period comets. This idea dates back much further \citep[see e.g.][]{Newton1891}, with \cite{Everhart1972} suggesting that it originated with Laplace. These `new' short-period comets can then undergo repeated interactions with Jupiter to randomise their orbital elements, though this process cannot explain all of the features in the short-period population \citep{Havnes1969, Havnes1970}. Whilst these studies were largely concerned with Jupiter mediating the transfer of long-period comets onto short-period orbits, the physics is very similar to that of capturing mildly hyperbolic ISOs that might encounter the Solar system.

`Oumuamua's discovery has led to a resurgence in studies of ISO capture. \cite{LingamLoeb2018} considered previous analytical studies to estimate the capture rate of interstellar comets by our Solar system. 
\cite{NamouniMorais2018} suggest that Jupiter's retrograde, co-orbital asteroid has an interstellar origin and was captured 4.5 Gyr ago. Integrating trajectories undergoing a close encounter with Jupiter, \cite{SirajLoeb2018} found that objects captured in this way may have orbits similar to those of the `Centaurs' with inclination $i>48^\circ$ and concluded that those may be captured ISOs.

\cite{Grishin2019} demonstrated that interstellar objects may be captured as a result of interactions with a primordial protoplanetary gas disc. Even though gas drag is an efficient way to capture ISOs, disc lifetimes are only $\sim10\,$Myr, about 500 times shorter than the age of the Solar system, implying that there was much more time for capture by purely gravitational dynamics. Nonetheless capturing ISOs into protoplanetary discs could be important for seeding planet formation \citep{Pfalzner2019}.

Knowledge of the density of ISOs is essential to understand the significance  of a potential captured population. Even more than 40 years before `Oumuamua's discovery, \cite{Whipple1975} was able to infer a density of interstellar comets of $10^{-3} \, \auq$. Shortly before `Oumuamua's Solar system fly-by, \cite{Engelhardt2017} used years of non-detections in three surveys to infer that comet-style ISOs of sizes $\ge1\,$km have density of at most $2.4\times10^{-4}\,\auq$, while asteroid-like ISOs (of the same size) could be 100 times more abundant. In the wake of `Oumuamua's discovery, \cite{Do2018} used its detection in 3.5 years of Pan-STARRS operation to put an upper limit of $0.2 \, \auq$ on the density of `Oumuamua-like ISOs, i.e.\ asteroids with size $\gtrsim100\,$m. These estimates are in broad agreement, given the difference in the assumed size of the objects.

Here, in the wake of the discovery of the first two interstellar objects, we revisit with $N$-body simulations the idea of ISO capture by the Sun-Jupiter system with a view to understanding the population that one might expect from repeated captures. 

\section{Numerical method}
\label{sec:numerics}
To model the encounter between each ISO and the Sun-Jupiter system, we use the time-reversible, adaptive $N$-body method described in \cite{Hands2019}. To leverage the parallelism in this code, we integrate the approach of $10^5$ ISOs per simulation and 404,800,000 in total. ISOs are modelled as test-particles, interacting only with the Sun and Jupiter, and remove them from the simulation if they come within the physical radius of either massive body.

\begin{figure*}
\begin{center}
\includegraphics[width=0.78\linewidth]{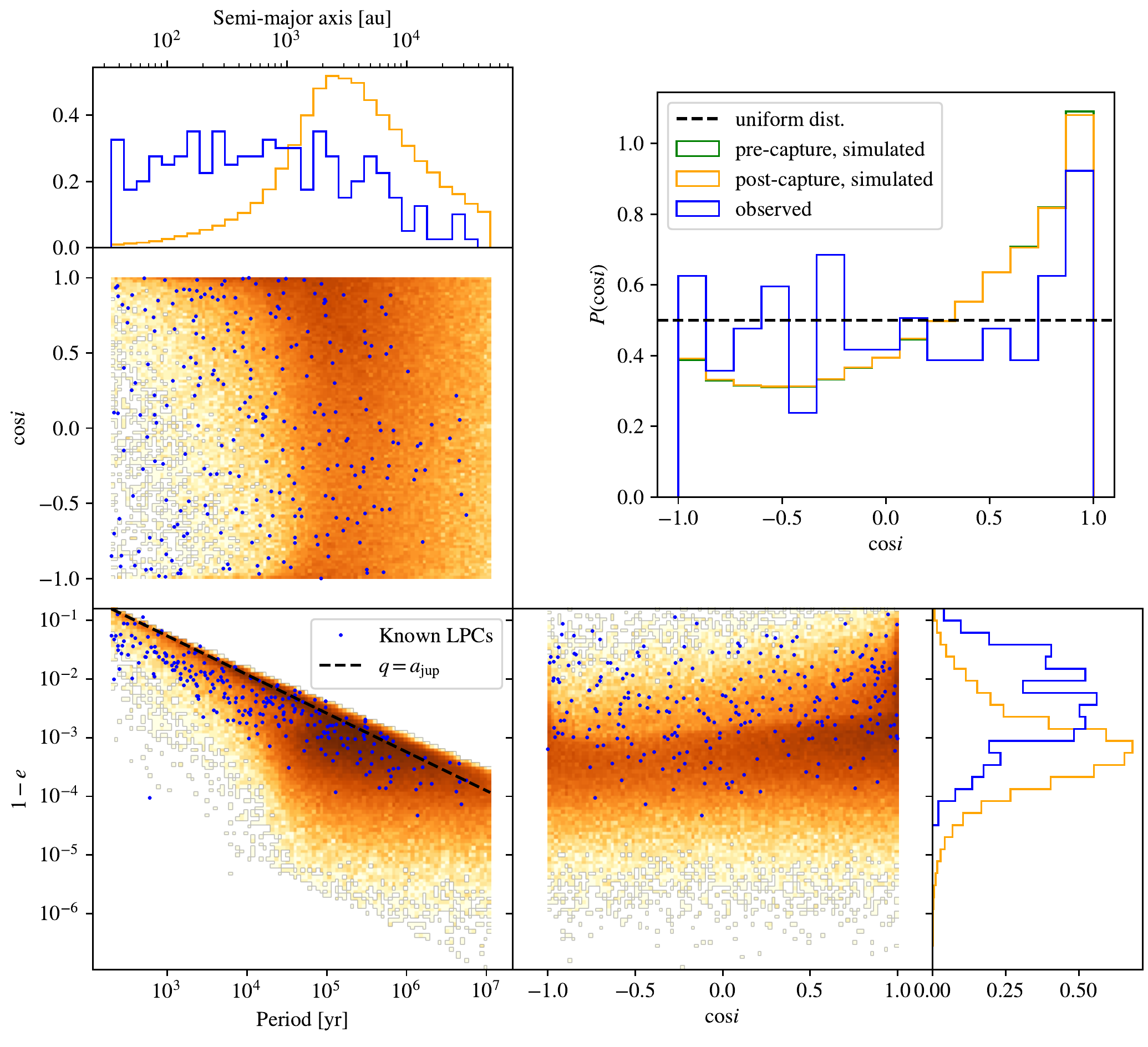}

\caption{Orbital elements for observed comets (blue) and captured objects from our simulations (orange). Based on their orbital elements, any known long-period comet could potentially be a captive. The black dashed line in the bottom left panel shows periapsides at the orbital radius of Jupiter. The top-right panel shows the distribution of cosines of inclination (with pre-capture values of eventual captives shown in green). Captured objects prefer to co-rotate with Jupiter, because such orbits tend to have smaller velocities relative to the planet and hence longer interaction times.
\label{fig:money_plot} }
  \end{center}
\end{figure*}

We begin with a Sun-Jupiter system in which Jupiter has a mass $M_{\mathrm{Jup}} = 9.546 \times 10^{-4} \, M_\odot$ and orbital elements $(a,e,i) = (5.202  \, \mathrm{au}, 0.048, 0^\circ)$. Throughout this work, inclination is measured relative to the $z$-axis in simulation coordinates, meaning the $x-y$ plane is the ecliptic, and all other orbital elements and phase-space coordinates are measured relative to the barycentre of the binary. Jupiter begins each run with a random mean anomaly $\ell$. Each ISO approach is characterised by its barycentric trajectory at $t=-\infty$,
\begin{align}
	\label{eq:comet:asymp}
	\vec{r}(t) = \vec{b} + t \vec{v}_\infty
\end{align}
with vectors $\vec{b}\perp\vec{v}_\infty$.
If $n$ is the number density of ISOs and $p(v_\infty)$ their speed distribution, the capture rate is
\begin{align}
    \label{eq:R}
	R &= n \int_0^\infty  v_\infty p(v_\infty)\,
	\mathrm{d} v_\infty
    \int \frac{\mathrm{d}^2\hat{\vec{v}}_\infty}{4\pi}
    \int \mathrm{d}^2\vec{b}
	\int_0^{2\pi} \frac{\mathrm{d}\ell}{2\pi}\;
	\quad \times
	\nonumber \\ &\phantom{=n}
	\Theta\left(-\frac{G(M_\odot+M_{\mathrm{Jup}})}{a_{\mathrm{lim}}}-E_{\mathrm{final}}\right)
\end{align}
with $\Theta$ the Heaviside function. The extra factor of $v_\infty$ in the integral accounts for the flux. Only simulated ISOs with semi-major axis $0<a\le a_{\mathrm{lim}}=50,000\,$au at the end of the simulation are considered captured, since orbits with larger $a$ are liable to be stripped by passing stars.

Assuming the ISOs follow the speed distribution of stars relative to the Sun, we model $p(v_\infty)=4\pi F(\vec{v}_\odot) v_\infty^2$, which is the natural low-$v_\infty$ limit for any smooth distribution $F(\vec{v})$ of space-velocities. For stars within 100\,pc, Gaia DR2 kinematic data \citep[e.g.][]{Schoenrich2019} indicate $F(\vec{v}_\odot)\approx5.5\times10^{-6}(\kms)^{-3}$ and that $p(v_\infty)\propto v_\infty^2$ holds up to $12\,\kms$. In our simulations we also limit the incoming ISO speed to $v_\infty\le v_{\infty,\max}=12\,\kms$, which makes little difference to our results since the chances of capturing such energetic objects are vanishingly low.

In order to reduce the computational costs and since orbits with large $b$ are less likely to be captured, in particular at high $v_\infty$, we restrict the maximum impact parameter to
\begin{equation}
    \label{eq:bmax}
    b \le b_{\max} =
    B/v_\infty
    \quad\text{with}\quad
    B\equiv 180\,\mathrm{au}\,\kms.
\end{equation}
In the limit of low $v_\infty$ this criterion allows initial hyperbolic orbits with periapse $q < 18.2 \,\mathrm{au}$, whilst at $v_\infty = 12 \kms$ orbits with $q < 10.0 \,\mathrm{au}$ are permitted. In all our integrations no initial condition with $b>0.9 b_{\max}$ resulted in capture, implying that the fraction of captures missed by the choice~\eqref{eq:bmax}, if any, is negligible. We also note that orbits with periapsides $q\gtrsim8\,$au are likely to come under the influence of the outer planets of the Solar system, which we do not model here. With these specifications, the Monte-Carlo sampling is uniform in $v_\infty^2$ and $(bv_\infty)^2$ between 0 and upper limits $v_{\infty,\max}^2$ and $B^2$, respectively, and the Monte-Carlo estimate for the volume capture rate, or \emph{rate function} \citep{Heggie1975}, becomes
\begin{align}
    \label{eq:mccaprate}
    Q \equiv \frac{R}{n} &= 
    2\pi^2 F(\vec{v}_\odot)\, v_{\infty,\max}^2B^2  \frac{N_{\mathrm{captured}}
    }{N_{\mathrm{simulated}}}
    .
\end{align}

In order to generate the initial positions and velocities for the integrations, we analytically evolve the hyperbolic barycentric orbit with asymptote~\eqref{eq:comet:asymp} to barycentric radius 800\,au. This is done for each of the $10^5$ ISOs in each simulation such that they initially form a spherical shell of radius 800\,au around the Sun/Jupiter system, with inclinations uniformly distributed in $\cos i$. We run each simulation for 2000\,yr, ensuring that even the lowest velocity ISOs have time to complete their passage of the Solar system.

\section{Results}
In the following we compare results from our capture simulations to known LPCs, whose data we take from the IAU Minor Planet Center%
\footnote{https://minorplanetcenter.net/data}. We therefore only consider simulated captives with periods $200\;\mathrm{yr}<P<$11\,Myr (corresponding to $a=50,000\,$au as mentioned in Section~\ref{sec:numerics}). From our 404.8 million orbital integrations, 192533 capture events meet these requirements, and a further 965 have $P<200\,$yr. From equation~\eqref{eq:mccaprate} this implies a rate function of
\begin{align}
    \label{eq:Q:result}
    Q \approx 0.051\,\auq \mathrm{yr}^{-1}
\end{align}
for $P>200\,$yr and 200 times smaller for $P<200\,$yr. The uncertainty in this number is of order few per cent and dominated by that for $F(\vec{v}_\odot)$. Combined with the density estimates in section 1, this implies a capture rate of $\simeq 12 \, \mathrm{Myr}^{-1}$ for $\gtrsim 1$ km sized comets, $\simeq 1200 \, \mathrm{Myr}^{-1}$ for $\gtrsim 1$ km sized asteroids and $\simeq 10^{4} \, \mathrm{Myr}^{-1}$ for `Oumuamua-like rocks.

Figure \ref{fig:money_plot} displays the orbital elements of these simulated captured ISOs and of known long-period comets for comparison. The capture mechanism produces comets with both longer periods and higher $e$ (lower $1-e$) than are observed, though all known long-period comets fit in the parameter space populated by the simulated captives. This raises the interesting possibility that interstellar objects might be hiding in plain sight in the LPC population. The figure also shows the distributions of $\cos i$ before and after capture. These distributions are very similar, indicating that the capture process makes very little difference to the inclinations, because for most captured objects a small change to their orbit suffices to bind them to the Solar system. There is an evident bias towards capturing objects on aligned pro-grade orbits ($\cos i\simeq1$), and a smaller but still clear bias toward capturing objects on anti-aligned (retro-grade) orbits ($\cos i\simeq-1$). This is simply the result of objects in the same plane as Jupiter having more opportunities to encounter the giant on their passage through the Solar system. Notably, these extra captives at $|\cos i|\simeq1$ tend to have smaller and less eccentric orbits than the remaining objects. Hence, captured ISOs may be most likely found in the ecliptic.
In fact, the preference for $\cos i\simeq1$ is also present in the comet data, although to a lesser degree. This may indicate that some of them are captives, but it may also, as for short-period comets, reflect an origin in a prograde disc of planetesimals. Note that these results contrast with those of \cite{SirajLoeb2018}, whose captives are of shorter $P$, lower $e$ and higher $i$. Cutting our results at their 75th percentile for semi-major axis ($a < 16$ au) shows that these Centaur-like objects form only 0.2\% of the overall captive population, though we do not see the same proclivity for high $i$ in this regime as they do.

\begin{figure}
\begin{center}
\includegraphics[width=0.98\linewidth]{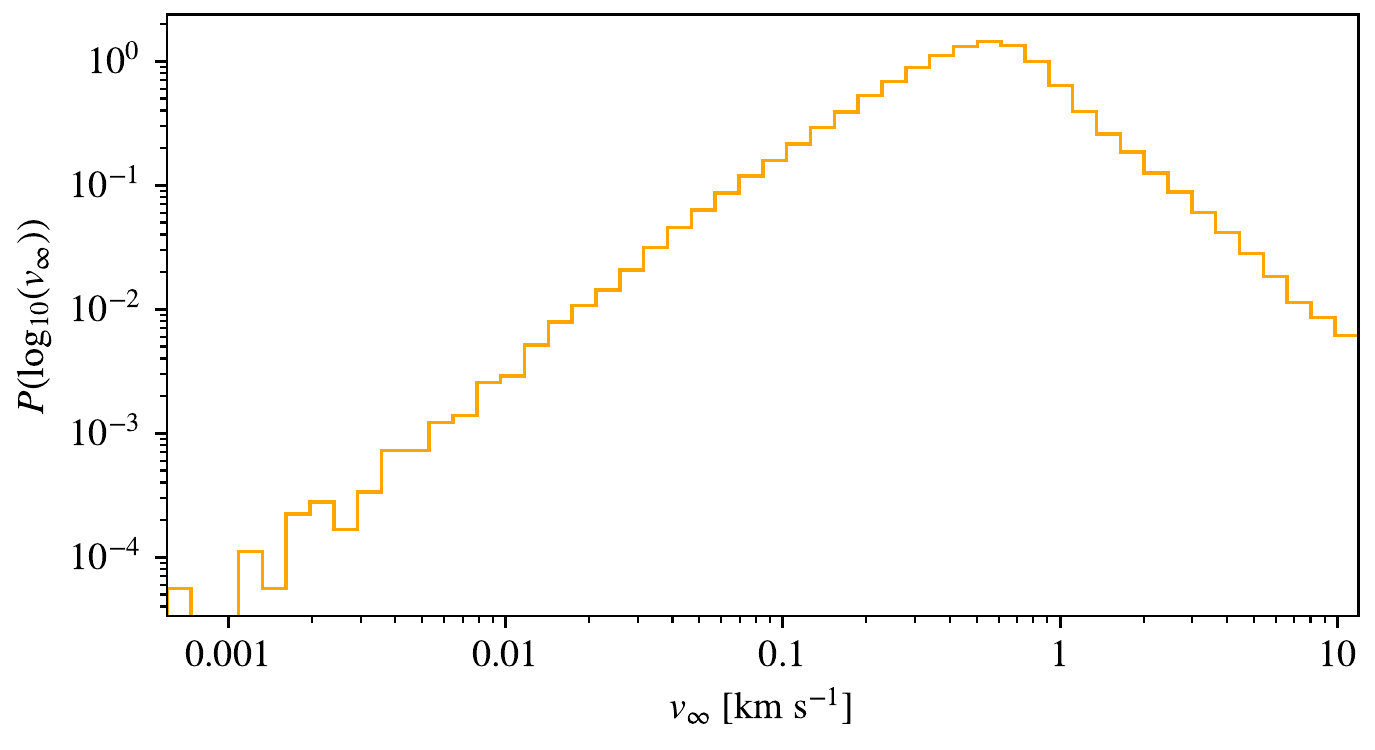}

\caption{Hyperbolic excess velocities on approach to the Solar system for ISOs that were captured in our simulations. 
The capture probability declines steadily towards 12km/s, leading us to believe that cutting the initial velocity distribution at 12 km/s has made little difference to the results. The preference for capturing low velocity ($v_\infty \simeq 0.6 \kms$) objects suggests that capture may only occur for a small fraction of ISOs. \label{fig:vinf} }
  \end{center}
\end{figure}

Figure \ref{fig:vinf} shows the hyperbolic excess speeds $v_\infty$ of all objects that were consequently captured, which are much lower than the 26\,\kms\ of `Oumuamua or 32\,\kms of Borisov. Indeed, velocities as high as these are common in the Solar neighbourhood, and hence captures are by implication rare as already pointed out by \cite{Valtonen1982}. The distribution of $v_\infty$ for captured objects follows the theoretical expectation and peaks near $(M_{\mathrm{Jup}}/M_\odot)^{1/2}$ times the orbital speed of Jupiter (Dehnen \& Hands in preparation).

\begin{figure}
\begin{center}
\includegraphics[width=0.98\linewidth]{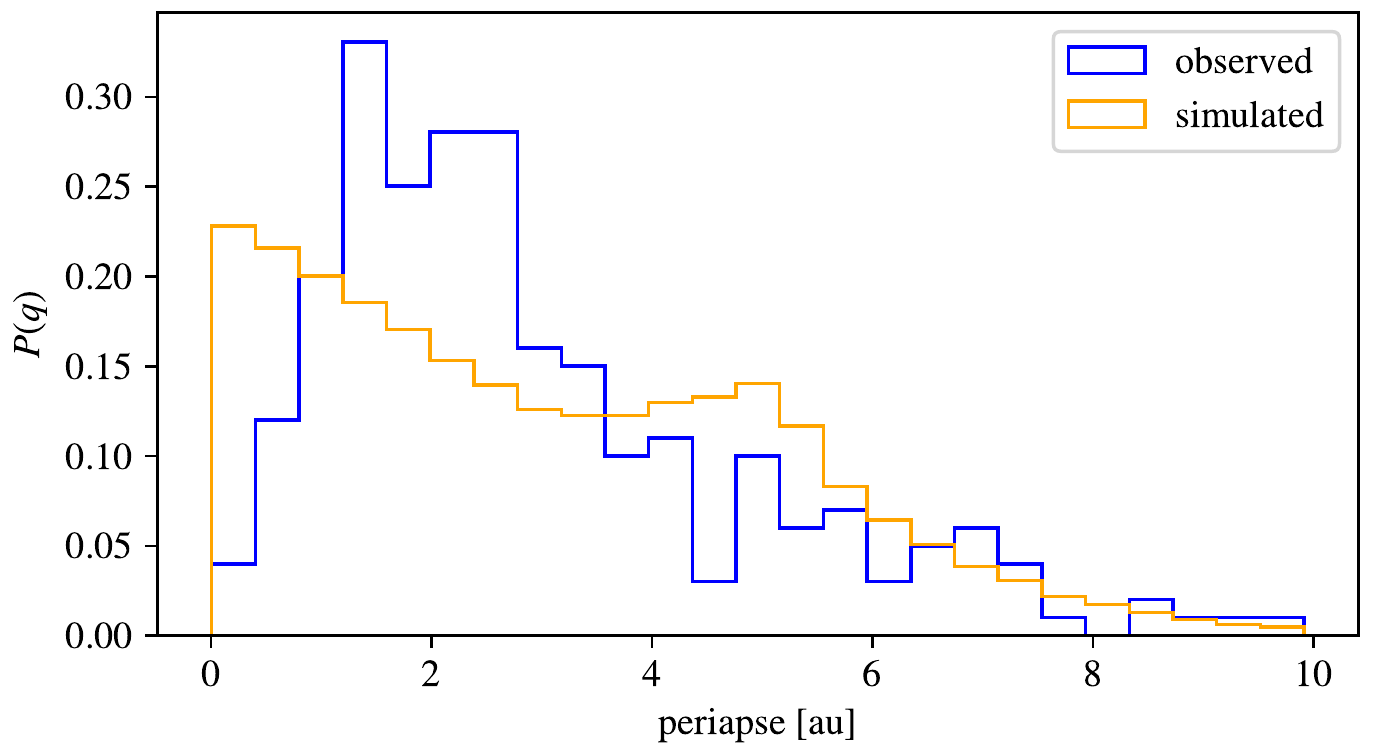}

\caption{Pericentres for known long-period comets in the Solar system and captured ISOs in our simulations. The two spikes at $q < 0.5 \mathrm{au}$ and  $q \simeq 5.2 \mathrm{au}$ may be the best places to look for captured ISOs. \label{fig:peri} }
  \end{center}
\end{figure}

Finally, Fig.~\ref{fig:peri} plots the perihelia $q$ of captured objects and observed LPCs (some caution should be exercised when comparing these distributions because of the unavoidable detection bias inherent in the LPC data). There is a strong preference for captives to be placed on orbits with perihelia at $q < 0.5\,$au, a region of parameter space in which very few LPCs have been found. We therefore tentatively suggest that future discoveries of LPCs with $q < 0.5\,$au are ideal candidates to be captured ISOs. A further spike is observed near 5.2\,au. This is a natural result of objects being captured by Jupiter, and suggests that comets with perihelia near Jupiter may be the best place to start looking for captured interstellar intruders, though one must be careful to distinguish these from native comets that have encountered the planet.

\label{sec:results}

\section{Discussion} \label{sec:discussion}

Our Oort cloud might be populated partially by objects that were directly captured from another Oort cloud around a passing star \citep{Hanse2018} or with somewhat greater chance from a planetesimal disc around a Solar sibling when the Sun was still in its birth cluster $\sim4\,$Gyr ago \citep{Hands2019}. Our results present a another potential origin for such alien comets: capture from inter-stellar space. Of course, these captured ISOs must have been liberated from their parent star by post-main-sequence evolution of planetary systems \citep{Hansen2017}, interactions with planets within their natal host system \citep{Raymond2018b,Raymond2018}, or close encounters in the stellar birth cluster \citep{Hands2019}, see also \cite{PortegiesZwart2018}.

\subsection{Detectability of alien stowaways}
The similarity between the orbital elements of known LPCs and our simulated captive ISOs raises the question of how one might distinguish between native and captive comet populations. Our simulated ISOs could be both volatile-rich, comet-like objects (such as 2I/Borisov) or volatile-poor rocky bodies (such as 1I/`Oumuamua). Although asteroids are thought to form and be ejected in lower numbers than comets \citep{Raymond2018b}, either breed of ISO might be captured by the Solar system onto LPC-like orbits. If the native LPCs all formed from the same reservoir of volatile-rich material, volatile-poor captives would stand out. 

We note that there is already a known population of 
non-cometary objects on comet-like orbits \citep[see e.g.][]{Fernandez2005}, the classic example being 1996 PW \citep{Williams1996}. Based on this object's dynamical history, \cite{Weissman1997} conclude that it is equally as likely to be an asteroid as an extinct comet, and suggest that such asteroids could have been emplaced into the Oort cloud early in the Solar system's life, or recently scattered from the inner Solar system onto eccentric orbits. Ideally of course, one wants to distinguish between asteroids and extinct comets observationally. Spectroscopy may make this possible, as demonstrated by the so-called `Manx' family of coma-less comets and in particular C/2014 S3. This object is on a classic LPC orbit, but displays little cometary activity, with spectroscopy revealing physical similarity to main-belt asteroids \citep{Meech2016}. As with 1996 PW, this result led to speculation that inner Solar system material may have been injected into the Oort cloud, though the absence of cometary activity is also consistent with this object being captured. Further modelling of both scenarios is required in order to ascertain if asteroid-like long-period objects are predominantly of extra-Solar origin. One potential sign of interstellar material is an unusual volatile composition, such as that of the CO-dominated C/2016 R2 \citep{Cochran2018,Mckay2019}. In such cases, different formation locations and dynamical as well as chemical histories of a comet's natal proto-planetary disc could explain compositional differences relative to native comets.

\subsection{Long-term evolution and steady-state population}
Whilst we do not attempt here to model the long-term behaviour of captured ISOs in detail, their orbits are sufficiently similar to known LPCs that studies of these can help us understand how captives might evolve. Specifically, the aforementioned work by \cite{Havnes1969} and \cite{Everhart1972} shows that repeated interactions with Jupiter can place some LPCs on short-period orbits.
This suggests that captives could also be hiding in the short-period comet population, in greater numbers than the 0.5\% that our simulations suggest. Conversely, perturbations by passing stars near aphelion or by massive bodies in the outer Solar system \citep[scattered planetary embryos,][]{Silsbee2018} could raise the perihelion distances of the captured ISOs out of the influence of the giant planets, placing them in the more traditional Oort cloud \citep[see e.g.][]{Dones2004}, where they may lie dormant for millions of years.

The capture rate for ISOs is dominated by small $v_\infty$ impacts and much larger than the value of one per 60\,Myr for typical ISO speeds of $v_\infty=20\,\kms$ \citep{Torbett1986}. Assuming those numbers and a lifetime of 0.45\,Myr for short-period comets \citep*{Levison1994}, \cite*{Gaidos2017} concluded that a steady-state population of captives is unlikely. However, this conclusion cannot be upheld in view of realistic capture rates and the long periods of most captive's orbits.

If the typical exponential lifetime for captured ISOs due to ejection by the giant planets, but also other processes such as stellar perturbations, is $\tau$, then the population evolves according to $\dot{N}=R-N/\tau$ and approaches a steady-state value of $N=R\tau$. From simulations of LPCs crossing the orbit of Jupiter \cite*{Horner2010} found $\tau\sim10\,$Myr. With this figure and our estimates for the capture rates, we expect a steady-state population of $\sim100$ comets of sizes $\gtrsim1\,$km and $\sim10^5$ `Oumuamua-like rocks. For objects with $P\sim1\,$Myr, a lifetime of 10\,Myr corresponds to a chance $\sim10\%$ per period for ejection. However, for very long period orbits, we expect this chance to approach 50\%, since their binding energies are so small that any weak interaction with Jupiter may result in ejection. This gives more conservative estimates of $\sim20$ and $\sim20000$ for the steady-state population of comets and `Oumuamua-like objects, respectively. 

Capture rates for short-period objects in our models are a factor of $\sim200$ smaller than for LPCs, but still at least 3 times larger than those of \cite{Torbett1986}. Taking $\tau=0.45$Myr with our short-period capture rates suggests even the SPCs could hide a handful of captured `Oumuamua-like objects.

The fraction of these long- and short-period populations combined that at any given time is within 6\,au of the Sun is only $3.3 \times 10^{-4}$ and has median semi-major axis of 52.2\,au. This suggests that
the chances of detecting them are small but not hopeless.

\subsection{Caveats and Future work}
As we demonstrate in a forthcoming paper (Dehnen \& Hands, in preparation), the capture rate owed to Saturn is $\sim5$ times smaller than for Jupiter and becomes comparable only at $v_\infty\lesssim0.1\,\kms$, while those from Uranus and Neptune are much smaller yet. However, at $v_\infty\lesssim0.5\,\kms$ most captures are caused by passages of Jupiter well outside its Roche sphere, when the interactions with other planets may not be neglected. Future work should explore the importance of those planets as well as the interplay of their interaction with impacting ISOs, both in terms of capture rate and detectability. Our simulations miss potential captures at $v_\infty>12\kms$, but the small number of captures at high $v_\infty$ suggest these missed events have negligible impact on our results.


\section{Summary}
We have performed numerical experiments to demonstrate the capture of interstellar objects by the Sun-Jupiter system. Our results suggest that the capture probability is small and the process only feasible for objects with asymptotic heliocentric speed $\lesssim4\,\kms$. Assuming these objects have velocities similar to local stars, we find a rate of capture of objects with number density $n$ of $R\approx 0.05 n\,\mathrm{au}^3\mathrm{yr}^{-1}$. With the current still rather uncertain estimates for $n$, this implies between a few and thousands per Myr for 1km-sized comets and 100\,m-sized asteroids, respectively, and suggests steady-state populations of $\sim100$ comets and $\sim10^5$ `Oumuamua-like objects are present in the outer Solar system.

Captured objects are typically on orbits very similar to those of long-period comets that humanity has observed for centuries, suggesting that this small population is hiding in plain sight. Despite the similarity between the captured and primordial comet populations, our results provide some hints as to the most likely orbits in which to find captured objects -- the captives prefer the ecliptic and perihelia not much larger than Jupiter's orbit. Finally, we suggest that long-period objects with little cometary activity are the best place to begin searching for objects of interstellar origin.

\section*{Acknowledgements}
This research has made use of data and/or services provided by the International Astronomical Union's Minor Planet Center. The authors thank the `ISSI `Oumuamua team', Avi Mandell, and Anna Boehle for useful discussions, and the reviewer Simon Portegies Zwart for helpful comments
. This work has been carried out in the frame of the National Centre for Competence in Research `PlanetS' supported by the Swiss National Science Foundation (SNSF).

\bibliographystyle{mnras}
\bibliography{flyby}
\label{lastpage}
\end{document}